\documentclass[useAMS,usegraphicx,usenatbib]{mn2e}
\usepackage{aas_macros}
\usepackage{grffile}
\usepackage{amssymb,amsmath,amsfonts}
\usepackage{multirow}
\usepackage[dvips, bookmarks]{hyperref}

\usepackage{color}
\newcommand{\rev}[1]{{#1}}

\numberwithin{equation}{section}
\newcommand{\msun}{M$_\odot$}
\newcommand{\hinv}{h^{-1}}
\newcommand{\myplot}[1]{\includegraphics[width=8cm,height=8cm]{#1}}
\newcommand{\myplottwo}[2]{\myplot{#1}\myplot{#2}}

\begin{document}
\title[Hierarchical Bound-Tracing Algorithm]{Resolving Subhaloes' Lives with the Hierarchical Bound-Tracing
Algorithm}

\author[J. Han et al.]
{Jiaxin Han,$^{1,2,5}$\thanks{jxhan@shao.ac.cn} Y.P. Jing,$^{1}$ 
Huiyuan Wang,$^{3,4}$, and Wenting Wang,$^{1,2}$\\
$^1$Key Laboratory for Research in Galaxies and Cosmology,
Shanghai Astronomical Observatory, Shanghai 200030, China\\
$^2$Graduate School of the Chinese Academy of Sciences,
19A, Yuquan Road, Beijing, China\\
$^3$Key Laboratory for Research in Galaxies and Cosmology,
University of Science and Technology of China, Hefei, Anhui 230026, China\\
$^4$Department of Astronomy, University of Science and
Technology of China, Hefei, Anhui 230026, China\\
$^5$Institute of Computational Cosmology, Department of Physics,
University of Durham, Science Laboratories,\\ South Road, Durham DH1
3LE\\
}

\maketitle

\begin{abstract} We develop a new code, the Hierarchical Bound-Tracing (HBT for
short) code, to find and trace dark matter subhaloes in simulations based on the
merger hierarchy of dark matter haloes. Application of this code to a recent
benchmark test of finding subhaloes demonstrates that HBT stands as one of the
best codes to trace the evolutionary history of subhaloes. The success of the code
lies in its careful treatment of the complex physical processes associated with
the evolution of subhaloes and in its robust unbinding algorithm with an
adaptive source subhalo management. We keep a full record of the merger
hierarchy of haloes and subhaloes, and allow growth of satellite subhaloes through
accretion from its ``satellite-of-satellites'', hence allowing
mergers among satellites. Local accretion of background mass is omitted, while
rebinding of stripped mass is allowed. The justification of these treatments
is provided by case studies of the lives of individual subhaloes and by the
success in finding the complete subhalo catalogue. We compare our result to 
other popular subhalo finders and show that HBT is
able to well resolve subhaloes in high density environment and keep strict
physical track of subhaloes' merger history. This code is fully parallelized and 
freely available upon request to the authors.
\end{abstract}


\section{Introduction}
In the hierarchical universe, cold dark matter haloes grow mainly through
mergers with surrounding smaller haloes. After the merger, the imprints of
progenitor haloes are not wiped out but in fact they can survive for quite a
long time as self-bound substructures called subhaloes\citep{Moore98, SKID,
Klypin99, Moore99}. Galaxies form inside dark matter haloes and can be traced by
dark matter subhaloes after the merger. Because the non-linear growth of
structure in the dark matter component can be well produced in N-body
simulations, it has become a standard approach to build galaxy formation models
on top of the dark matter halo merger history \citep[see e.g,][for recent
reviews]{Baugh06,Benson10}. Constructing the full hierarchy of the merger
history requires the identification and linking of dark matter subhaloes
across cosmic time. Many algorithms have been developed to accomplish this
job, all based on some of the following characteristics of a subhalo:
\begin{enumerate}
\item It is an overdense region inside its host halo\label{c_o}
\item It is self-bound so that it is dynamically significant\label{c_b}
\item It was a halo before it merges into its current host halo\label{c_h}
\end{enumerate}

Most subhalo finders utilize only contemporary particle distribution
and focus on the first two characteristics, while the merger tree is
constructed by a subsequent matching of subhaloes at different
epochs. For example, the Hierarchical Friends-of-Friends \citep[HFoF;][]{Klypin99}) 
algorithm, which is an extension to the standard
Friends-of-Friends \citep[FoF;][]{fof} halo finder with multiple
resolutions, makes use of the first characteristic only. AMIGA Halo
Finder \citep[AHF;][]{AHF}, SUBFIND \citep{SubFind} and
SKID \citep{SKID}, which collect local overdense particles and then
eliminate unbound particles, are based on the first two
characteristics. Although these percolation based algorithms are able
to find subhaloes using a single simulation output, a strong resolution
problem can arise in the central part of the host halo due to
ambiguity of separating member particles of a subhalo from the
background ones in the high density region \citep{MHT,Muldrew10}. In
fact in a recent extensive halo-finder comparison project \citep{Mad},
those participating finders based on configuration information only
all struggle to recover substructures in the central high-density
region of a host halo (see also \ref{s:MAD}). One way to improve the
resolution of subhaloes in the high density region is to use six
dimensional phase space information, as done in, e.g, the Hierarchical
Structure Finder \citep[HSF;][]{HSF}) and ROCKSTAR \citep{rockstar}.
Another way out of this problem is to appeal to the third
characteristic and utilize the evolutionary history of
subhaloes. Because subhaloes are remnants of dark matter haloes, they can
be identified by tracing the member particles of their progenitor
haloes. The first attempt along this direction was \citet{Tormen97} who
just considered the third characteristic. Later in \citet{Tormen98}
self-boundness was also added to define a subhalo. Only those haloes
which fall directly into the final halo of interest were examined
until the work of \citet{G10}(hereafter G10) where they extended their
SURV code \citep{Tormen04,G08} to include subhaloes inside subhaloes. The
same method was also implemented in the 'MLAPM halo tracker' \citep[MHT;][]{MHT}.

While simple as the idea looks, the subhalo identification through tracing a
merger history still faces several difficulties in practice. The most
challenging aspect of the problem is how to trace subhaloes robustly over
several orbital periods. Mass loss is the primary process associated with the
subhalo evolution due to gravitational striping and harassment. In such a
simplified picture, once a particle becomes unbound to a subhalo, it no longer
needs to be traced any more. However, as we will show, this kind of successive
tracing is dangerous since artificial loss of bound particles can accumulate
through every tracing step, eventually triggering a runaway loss of a subhalo's
bound particles. On the other hand, if one always traces all the particles from
a subhalo's progenitor halo, a straight-forward unbinding algorithm may also
fail to find a self-bound structure once the subhalo has been substantially
stripped compared to its progenitor halo, due to the large amount of unbound
particles.  Besides the stringent requirement on the robustness of tracing,
second order effects can lead to mass growth for satellite subhaloes. As we will
show, both accretion and merger can continue to happen for satellites even
within the virial radius of the host halo. Ignoring these process will
under-estimate the mass of traced subhaloes.

In this work we present a new tracing code (Hierarchical
Bound-Tracing, HBT hereafter) to identify subhaloes and construct the
merger tree.  The key to our code is a robust unbinding algorithm
together with adaptive source-subhalo management. With these recipes we are
able to walk through the cosmic age to capture every subhalo ever
alive, as will be described in section~\ref{s:algo}. Our careful
tracing of subhaloes' hierarchy enables us to apply the unbinding
algorithm recursively, naturally allowing satellite accretion and
merger. The success of HBT is demonstrated in section~\ref{s:result}
to have high completeness. Through comparison to a configuration space
subhalo finder we also show that HBT subhaloes are in general more
robust and more extended. The results are summarized in
section~\ref{s:conc}. HBT's implication for galaxy formation models
and prospects are discussed in section~\ref{s:discussion}. Two
concordence LCDM simulations, \rev{both with cosmological parameters $\Omega_M=0.3$, $\Omega_\Lambda=0.7$, and $\sigma_8=0.9$,} are used for the tests and comparisons in
this work; one is a zoomed-in re-simulation of a $10^{15}\hinv$\msun~ 
galaxy cluster with a particle mass $m_p=1\times10^8\hinv$\msun~ 
carried out with GADGET II(\cite{GADGET2}), and the other is a
cosmological simulation with a boxsize $100 h^{-1}\mathrm{Mpc}$ using
$512^3$ particles\citep{Jing02}.

Our algorithm also makes use of characteristics \ref{c_b} and
\ref{c_h} and it turns out that the first characteristic is
automatically satisfied. The difference between our algorithm and that
of G10 is mostly technical, mainly in the time direction of tracing a
subhalo: HBT proceeds from the earliest epoch, naturally finding the
full hierarchy of subhaloes together with extracting the merger tree in
only one walk through the cosmic time; while G10 tries to figure out
the subhalo hierarchy level by level by subsequently revisiting
earlier snapshots. The MHT code depends on the 'MLAPM halo finder'
\citep[MHF;][]{MHT} to first identify haloes as well as subhaloes from
the halo formation time. Then the haloes and subhaloes are tracked in
subsequent outputs. In HBT the growth, merger, and stripping of haloes
and subhaloes are handled in a unified way starting from the earliest
resolved haloes, and the full hierarchy of the subhalo merger tree is
resolved. HBT also stands out in the sense that it is the first time
that various systematic issues in a tracing algorithm have been
investigated and the tracing results have been carefully assessed.

\section{Algorithm}\label{s:algo} 

\subsection{Overall tracing algorithm} 

HBT starts the tracing of subhaloes from input halo catalogues for a
sequence of simulation outputs.  \footnote{Usually 60 snapshots for a
  LCDM simulation starting from an initial redshift where the first
  several haloes can be found, e.g. $z_{ini}\sim 20$, is sufficient to achieve high
  completeness in the present-day subhalo catalogue.  Interested
  readers can refer to Appendix~\ref{s:time_resolution} where we
  investigate how the algorithm depends on the time resolution of
  simulation outputs.} In the current implementation we adopt the
simple and widely-used Friends-of-Friends (FoF) halo finder to
construct the input catalogues, because it has been demonstrated that
virialised haloes and subhaloes are included in these FoF haloes
(Springel et al. 2001), and also because it does not cut subhaloes near
halo boundaries as, for example, an spherical-overdensity halo finder
(\cite{LC94}) would do. For our two LCDM simulations, the nominal
linking-length $b=0.2$ is adopted. With halo catalogues in hand, the
HBT algorithm can be summarized in one sentence: HBT builds and
traverses the halo merger tree and finds the self-bound structure for
every halo at every snapshot after its birth.

Specifically, the subhalo identification breaks into two steps: first
selecting candidate particles which contain the members of the
self-bound structure, and then removing the unbound ones. We call the
initial collection of the candidate particles, out of which the
self-bound part gives rise to a subhalo, a \emph{source-subhalo} (or
source for short).  In HBT we work with two types of subhaloes in each
halo: a central subhalo which is the dominating subhalo within a host
FoF halo, and satellite subhaloes which are the remaining ones if any.
Starting from the highest redshift, haloes without progenitors are fed
to an unbinding procedure as sources of their central subhaloes. At the
next snapshot, the particles of these source subhaloes are tracked to
identify their host haloes. If more than one source is found to
reside in the same host, a halo merger is identified.\footnote{It also
  happens that a source's particles are distributed into multiple host
  haloes. In this case the source is split as described in
  Appendix~\ref{ss:split}.} The progenitor sources are then unbound
to select the most massive self-bound structure as the central
subhalo, and others as satellites.  The sources for the central
subhaloes are updated to be the current host halo excluding particles
from any self-bound satellites, and unbound again to allow for growth
of centrals. In this kind of tracing process, haloes without
progenitors create new branches in the merger tree and give birth to
new centrals, while mergers connect branches and transform centrals
to satellites.

Since the particles of satellite subhaloes are traced from their progenitors,
this is equivalent to assuming that no accretion of host halo's particles can
happen for satellite subhaloes. In Appendix~\ref{s:acc} we explicitly test this
assumption and show that local accretion has little effect on the tracing in
the long term.

We record the progenitor-descendant information as we proceed. This way we get
the subhalo catalogues and the merger history of these subhaloes at the same time.
    \begin{figure*}
    \includegraphics[width=\textwidth]{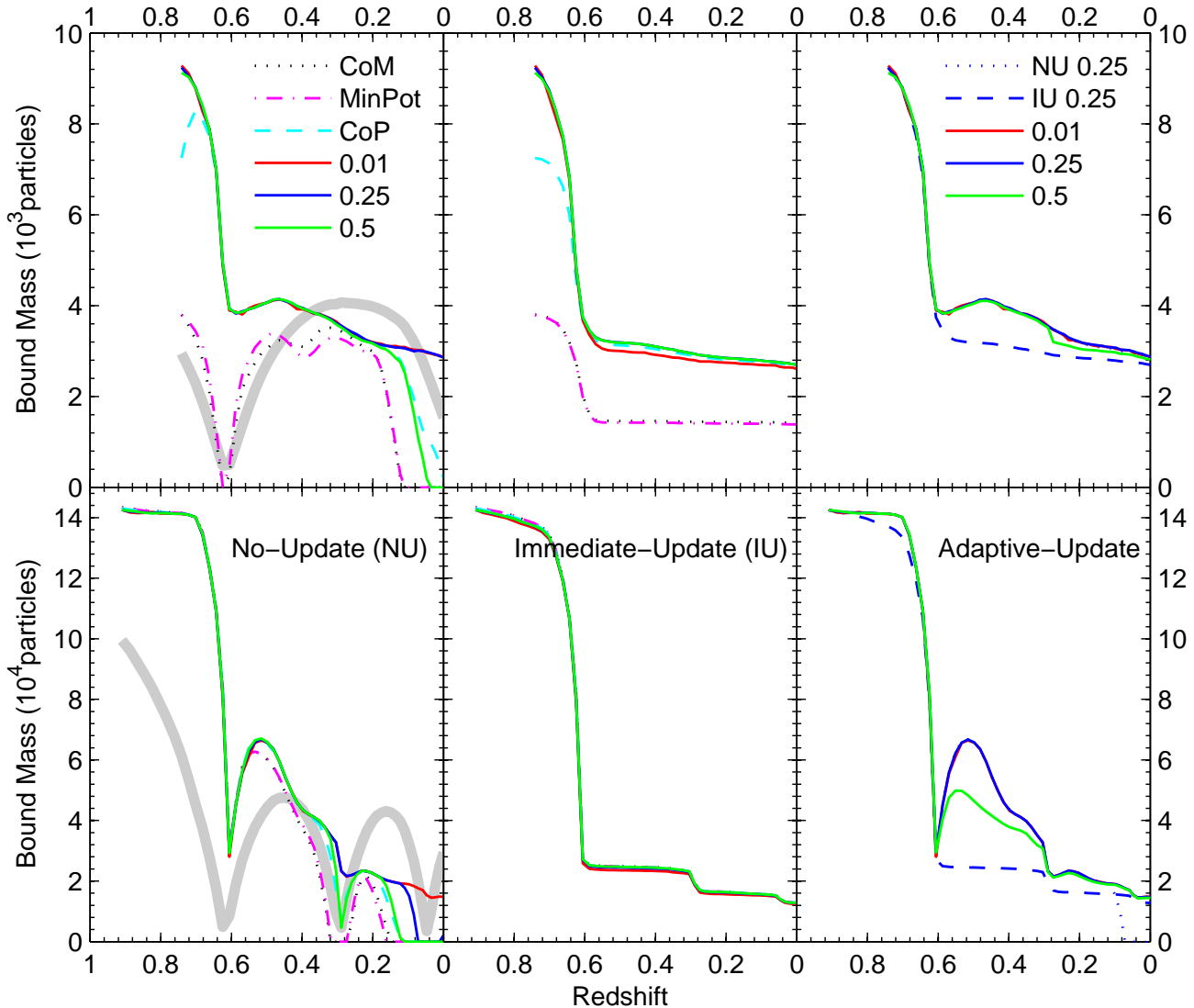}
    \caption{Robustness of various unbinding algorithms. Left: Subhalo
      mass evolution calculated when the progenitor halo is used as a
      source without any updates during tracing. The solid lines are
      results of core-averaged unbinding, with different
      $\mathtt{CoreFrac}$ as given in the legend. The other three
      lines denote the results when three commonly-used reference
      frames are adopted; the dotted line is for the centre of
      Mass(CoM) frame, the dot-dashed line for the MinPot frame, and
      the dashed line for the CoP frame.  These reference frames are
      defined in the text. The thick grey line shows the halo-centric
      distance of the subhalo, in an arbitrary unit. Middle: The same
      as in the left column, but the subhalo at last tracing
      step is used as a source. Right: Subhalo mass evolution calculated when a
      self-adaptive source is used. The solid lines are the results of
      self-adaptive core-averaged unbinding with different
      $\mathtt{CoreFrac0}$ parameters.  Over-plotted are the results
      of the core-averaged unbinding with no source update (the dotted
      line, overlapping with the solid blue line in top panel; as in
      the left column) and with immediate updates (the dashed line; as
      in the middle column), with $\mathtt{CoreFrac}=0.25$. Top panels
      are for halo S51G86, while bottom panels for halo
      S43G11. }\label{f:robustness} \end{figure*}
  
\subsection{Tracing Satellites Robustly and Efficiently}\label{s:robust}

As we have mentioned before, it is the most challenging part for a tracing algorithm to trace satellite subhaloes robustly and
efficiently. This is a
problem mostly because unbinding can be fastidious about the quality of source
subhaloes. Either too big or too small a source subhalo can lead to failure in
unbinding. This can be understood as follows. The definition of boundness
depends on the reference frame in which to calculate the kinetic energy and on
the assembly of particles from which the potential energy is obtained. For a
self-bound subhalo, the reference frame is defined to be at rest with respect
to a certain ``centre'' of the subhalo (e.g, centre of mass) which removes the
bulk motion of the system, and the potential energy is summed over all the
particles in the subhalo. Starting from a source subhalo with bound and unbound
particles, one still seeks a centre which is roughly at rest with the final
one, and uses this frame to calculate the binding energy. An overly large source
subhalo with too many unbound particles would probably give a centre which
deviates too much from the underlying subhalo, while an overly small one
consisting of only a small portion of the real subhalo would give too shallow
potential, both yielding severely biased result or even completely missing the
subhalo. For example, as shown in \cite{Hayashi03} an NFW halo truncated to a
radius of $<0.77$ times the scale radius is completely unbound.  Thus a source
subhalo is expected to be a slightly enlarged assembly of particles containing
the final subhalo. In addition, the unbinding procedure is required to be
robust, especially if the source is strongly contaminated. We make efforts in
both aspects to improve the robustness of HBT.


 In HBT we implement a core-averaged unbinding algorithm designed to tolerate contamination. Starting from a source
subhalo, unbound particles are removed iteratively till the bound mass
converges. For each iteration, the reference frame is chosen to be the centre
of mass and bulk velocity of an inner-most core consisting of a certain
fraction \texttt{CoreFrac} of the remaining particles with the
lowest-potential. This provides a robust estimate of the underlying true
reference frame in presence of contaminations.  More detailed description of
this algorithm can be found in Appendix~\ref{sec_unbind}. 


Besides improving unbinding, our HBT code updates satellite sources
adaptively, to keep the amount of contaminating particles under
control, while still maintaining a large enough reservoir of candidate
particles.  The idea is to reduce the size of the source
conservatively every time a subhalo has been stripped to a fraction
\texttt{CoreFrac0} of the current source mass $M_{src}$.  To ensure
that the new source with mass $M_{src2}$ is still larger than the
current subhalo, we replace the source with the subhalo's progenitor
found at the time it was first stripped to a fraction
$\sqrt{\mathtt{CoreFrac0}}$ of the original source mass.  With the new
source constructed this way, it is ready to see that the current
subhalo automatically becomes the new source in the next loop of
source updates, saving the trouble to search for new sources upon
every update.  The source for a subhalo before infall into a host halo
is simply its host halo with all the satellite particles masked out.
After infall this kind of source updates start to take effect all the
way along the subhalo's history.  Note that the removal of unbound
particles not only reduces contamination, but also reduces the amount
of calculation for unbinding.  With source subhaloes updated this way,
the unbinding procedure is always protected to work under the condition 
$\mathtt{CoreFrac0}\cdot M_{src}<M_{sub}<\sqrt{\mathtt{CoreFrac0}}\cdot{}M_{src}$,
substantially reducing the amount of work for unbinding and making the
tracing more robust.

The \texttt{CoreFrac} parameter adopted by the unbinding procedure is initially
\texttt{CoreFrac0}, and updated with
$\mathtt{CoreFrac}=M_{src2}/M_{src}\cdot{}\sqrt{\mathtt{CoreFrac0}}$ every time
we update the source. This is to ensure that the smallest subhalo out of
$M_{src}$ does not use an over-large core when
$M_{src2}<<M_{src}\cdot{}\sqrt{\mathtt{CoreFrac0}}$ due to discreteness in
simulation output time. In the current implementation, we adopt
$\mathtt{CoreFrac0}=0.25$. 

In the development stage we also tried several other reference frames
for unbinding, including: \begin{enumerate} \item centre of Mass (CoM)
  frame: Take the centre of mass of source particles as centre and the
  average velocity as the bulk motion.  \item centre of Potential
  (CoP) frame: Take the potential weighted centre of mass as centre
  and the potential weighted average velocity as the bulk
  motion.  \item Minimum Potential (MinPot) frame: Take the position
  of the particle with minimum potential energy as centre and average
  velocity of the source subhalo as the bulk motion. We avoid using
  the velocity of the minimum potential particle considering that it
  may have large dispersion with respect to the bulk
  motion.  \end{enumerate} For the source construction, we also
tried two simple strategies. One is to always use the source from
the infall without any updating (no-update), and the other is to update it
aggressively by using the progenitor subhalo at snapshot $n-1$ as the
source for a subhalo at snapshot $n$ (immediate-update).
       
  In Figure~\ref{f:robustness} we compare the performance of various
  combinations of these unbinding and source construction recipes, by
  applying them to two subhaloes in the cluster simulation.  Top panels
  show the case for tracing halo S51G86 (named after snapshot ID and
  halo ID just before the infall, with the merger mass ratio 0.0016
  and the initial orbital circularity 0.16).  The left column is the
  ``no-update'' regime, where the source is taken to be the progenitor
  halo at the infall, while the middle column does ``immediate update''
  and uses the subhalo at the previous tracing step as the source.  It
  can be seen that the CoM frame method and the MinPot frame method
  have similar performance; both methods fail to identify more than a
  half of the bound population. The performance is worse at a smaller
  halo-centric distance where the tidal force is stronger or when the
  bound part is smaller hence there are more contaminating particles.
  In general the core-averaged unbinding algorithm has the best
  performance, due to its ability to seek out a tight core to
  represent the majority of the bound particles. Still an
  over-estimation of the core size (e.g.  $\mathtt{CoreFrac}=0.5$
  after $z=0.2$ with no update) can cause an obvious drop in
  subsequent subhalo mass.  As shown in the top left panel, the
  subhalo gradually loses mass as it spirals into the centre of its
  host, reaches a minimum at pericentre passage near $z=0.6$, and then
  re-gains mass as it moves out. Those regained particles are also
  from the initial infalling halo, and we found that they are also
  mainly bound particles before the pericentre passage. This
  \emph{rebinding} of once unbound particles are suppressed in the
  top middle panel, where monotonic decrease in subhalo mass is forced
  by keeping only bound particles from last snapshot as the source. In
  an extreme case as shown in the bottom panels for halo S43G11(with
  merger mass ratio 0.02 and initial orbital circularity 0.10), the
  rebinding process can increase the subhalo's mass by a factor of
  2.3 after pericentre, and suppression of the re-capture can yield an
  under-estimation of subhalo mass by a factor of 2.8, comparing the
  left and middle panels. Right column shows the result when the
  adaptive update of source subhalo is applied. With the source size
  under control, the problem of an overly large core is
  avoided. Because $M_{src}$ marks the maximum subhalo mass allowed,
  $M_{src}<M_{sub}/{\mathtt{CoreFrac0}}$ shows that the maximum factor
  by which a satellite subhalo is allowed to grow through rebinding
  is $1/{\mathtt{CoreFrac0}}$. Thus a big $\mathtt{CoreFrac0}$ would
  still suppress mass growth, as in the case of the bottom right panel
  with $\mathtt{CoreFrac0}=0.5$.
  
\begin{figure}
\includegraphics[width=0.5\textwidth]{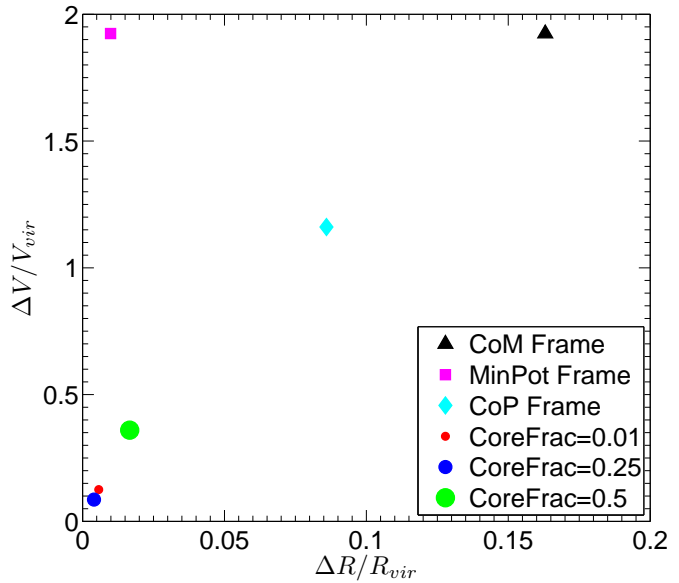} 
\caption{The reference frames in
    different unbinding algorithms for halo S51G86. For each algorithm, we plot
    the difference of the initial estimation of the origin and the relative
    velocity of the reference frame from the final reference frame found for
  the subhalo in the HBT code, normalized by the halo virial radius and virial
velocity.}\label{f:centre} 
\end{figure}

 Figure~\ref{f:centre} further shows the reason for the performances
 of different unbinding algorithms. At the time of infall, using the
 reference frame of the final self-bound subhalo in the
 $\mathtt{CoreFrac0}=0.25$ unbinding result as the standard one, we
 check the difference of the initial estimation of the reference
 frames in each algorithm from that standard frame, and plot them
 normalized by the host halo virial radius $R_{vir}$ and virial
 velocity $V_{vir}=\sqrt{G M_{vir}/R_{vir}}$.  The CoM, MinPot, and
 CoP frames all have large velocity residuals with respect to the
 bound structure, although the MinPot frame has a negligible
 positional displacement\footnote{Note that we also adopted the
   average velocity as reference for MinPot frame.}. However, the
 initial CoM frame also has a large positional displacement. The
 assignment of these displaced frames would inevitably result in the
 artificial loss of bound mass in these three algorithms.

The most-bound particle of an HBT subhalo is almost always located at
the centre of mass position of the core, with a displacement within
one softening length, reflecting the definition of our boundness.

\subsection{Allowing accretion and merger within satellite systems}  

Although the dominating physical process in satellite mass evolution is tidal
mass loss, it has been shown in \citet{Simha09} that mass accretion and merger
is not terminated for satellite galaxies. \citet{Angulo} also find that the
merger rate between satellite subhaloes and that between a satellite and a
central subhalo are comparable for satellites smaller than 0.01 times the host
halo mass, and that most of the satellite-satellite mergers happen between
subhaloes which were once in central-satellite relation. These satellites and
``satellite of satellite''s as well ``satellite of satellite of satellite''s
and so on define hierarchical satellite systems which could persist for a long
period up to many Gyrs as observed in \citet{White10}, and these systems are
places where satellite accretion and mergers can still happen. 

In HBT we keep record of this ``sat-of-sat'' hierarchy. Each central
subhalo is aware of its satellites, and each satellite is aware of its
sat-of-sats which it had accumulated before its infall and which are
still alive. There are occasions when a subhalo is ejected from a
host, which is quite common as shown in \citet{Lin03} 
\citep[see also][]{MHT3,Sales07,Ludlow09}. In this case the
ejected subhalo is also removed from the sat-of-sat list that contains it.
 Here we use the term ``sat-of-sat'' instead of
``sub-in-sub'' (see,e.g.,\citet{Aqua}) to emphasize that this relation
between subhaloes is a historical or dynamical relation, which may not
correspond to the spatial nesting relation in the latter case because
of separation of orbits due to host halo's tidal force or multiple
satellite interactions. We allow the accretion of mass within each
subhalo's sat-of-sats.  This is achieved by implementing the unbinding
procedure recursively: for a hierarchy of source subhaloes, starting
from the highest level source subhalo, we feed the unbinding procedure
with both particles from the current source subhalo and particles
which are removed from recursive unbinding of lower level source
subhaloes.\footnote{To avoid adding satellite particles multiple times,
  we define the source of a central subhalo to be the host halo
  excluding all the particles from its satellite \emph{sources} when
  recursive unbinding is applied.} This means particles removed from a
satellite can have a chance to be accreted into a higher level
subhalo.  When a satellite subhalo dies with the majority of its
particles accreted by a higher level satellite subhalo, a
satellite-satellite merger happens.

Ignoring the effect of satellite merger could result in loss of subhaloes during
tracing. For example, a subhalo with 7000 particles (or about 1/1000 the host
halo mass) is missing in the HBT result for the resimulation when satellite
merger is switched off. This is triggered by a satellite major merger event
between two satellites which were once in the central-satellite relation. After
merger the particle kinetic energies are increased.  If one still uses
particles from only one source subhalo to do the unbinding, the potential would
not be deep enough to bind the particles, resulting in the loss of the new
subhalo.

\begin{figure}
\includegraphics[width=0.5\textwidth]{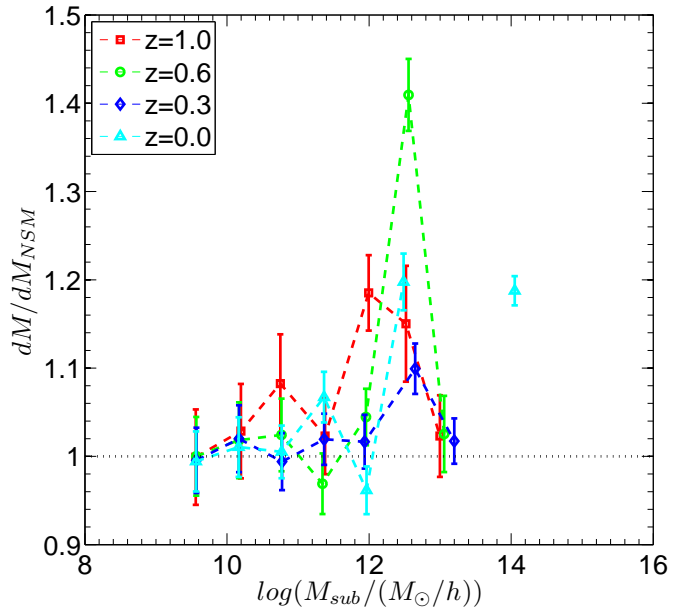}
\caption{The Effect of satellite merger. We plot the ratio of the
  total subhalo mass in each logarithmic mass bin between HBT results
  with ($dM$) and without ($dM_{NSM}$) satellite merger for the
  resimulated cluster. The errors are propagated from Poisson errors 
  in the total number of particles within each bin. 
  Different colour lines with symbols are for
  different redshifts, while the horizontal dotted line is the $1:1$
  reference.}\label{f:sat_acc}
\end{figure}

In Figure~\ref{f:sat_acc} we show the ratio of the total mass
contained in each subhalo mass bin for the resimulated cluster, found
by HBT with and without the satellite merger. This is equivalent to
the ratio of the mass weighted subhalo mass function $M_{sub}dN/d\ln
M_{sub}$\citep[see e.g.,][for more details of the mass
  function]{Gao04b}. Although allowing merger among satellite subhaloes
has almost no effect on small subhaloes, it can enhance the mass
function at the high mass end by 20 percent, a fraction close to the
fractional mass contribution from satellites to host haloes. The strong
fluctuations at the high mass end reflect the occurrence of significant
satellite merger events.

\section{Results and comparison}\label{s:result} To show that HBT has a
superior tracing ability to recover subhaloes in high density regions
and to compare with other subhalo finders, we have applied HBT to one
test case provided by the "haloes Gone Mad" project\citep{Mad}, to
examine a subhalo's mass evolution. To show that HBT
  completely recovers the subhalo population, we compare the subhalo
  mass functions from HBT for our simulations with those
  from SUBFIND, as well as with a fitting formula. The comparison
shows that HBT subhaloes are more massive than those from SUBFIND by 10
to 20 percent, a fact that is also observed when comparing the size of
subhaloes from the two codes. The reason for this difference is further
revealed in the density profile, where SUBFIND subhaloes show sharp
truncation near tidal radii while HBT subhaloes are more extended.  We
also show the one-to-one matching result between HBT and SUBFIND
catalogues. \rev{HBT has also been compared with many other subhalo finders in 
the subhalo-finder comparison project\citep{Notts}, and is found to have good performance.}

\begin{figure*}
\myplottwo{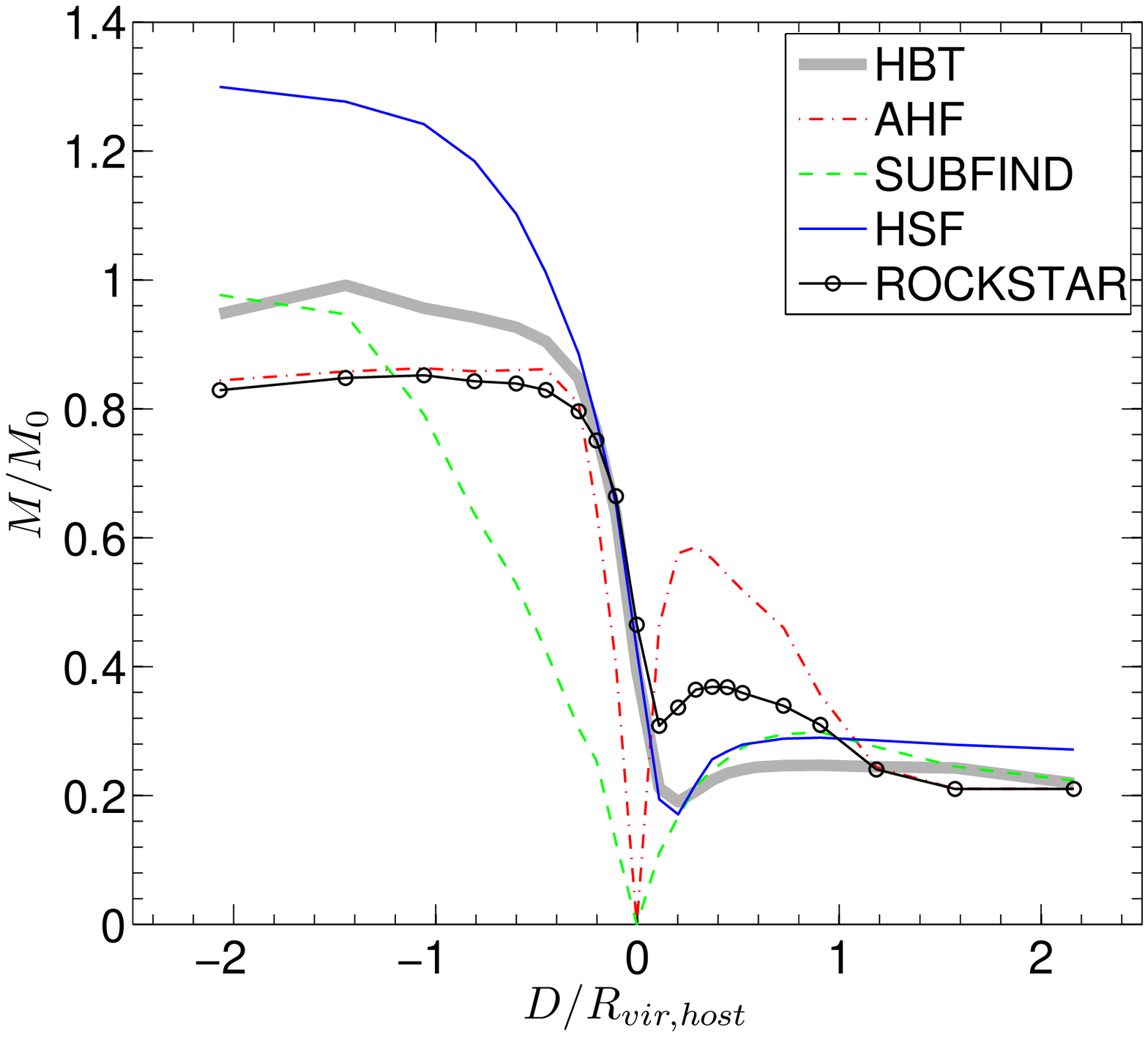}{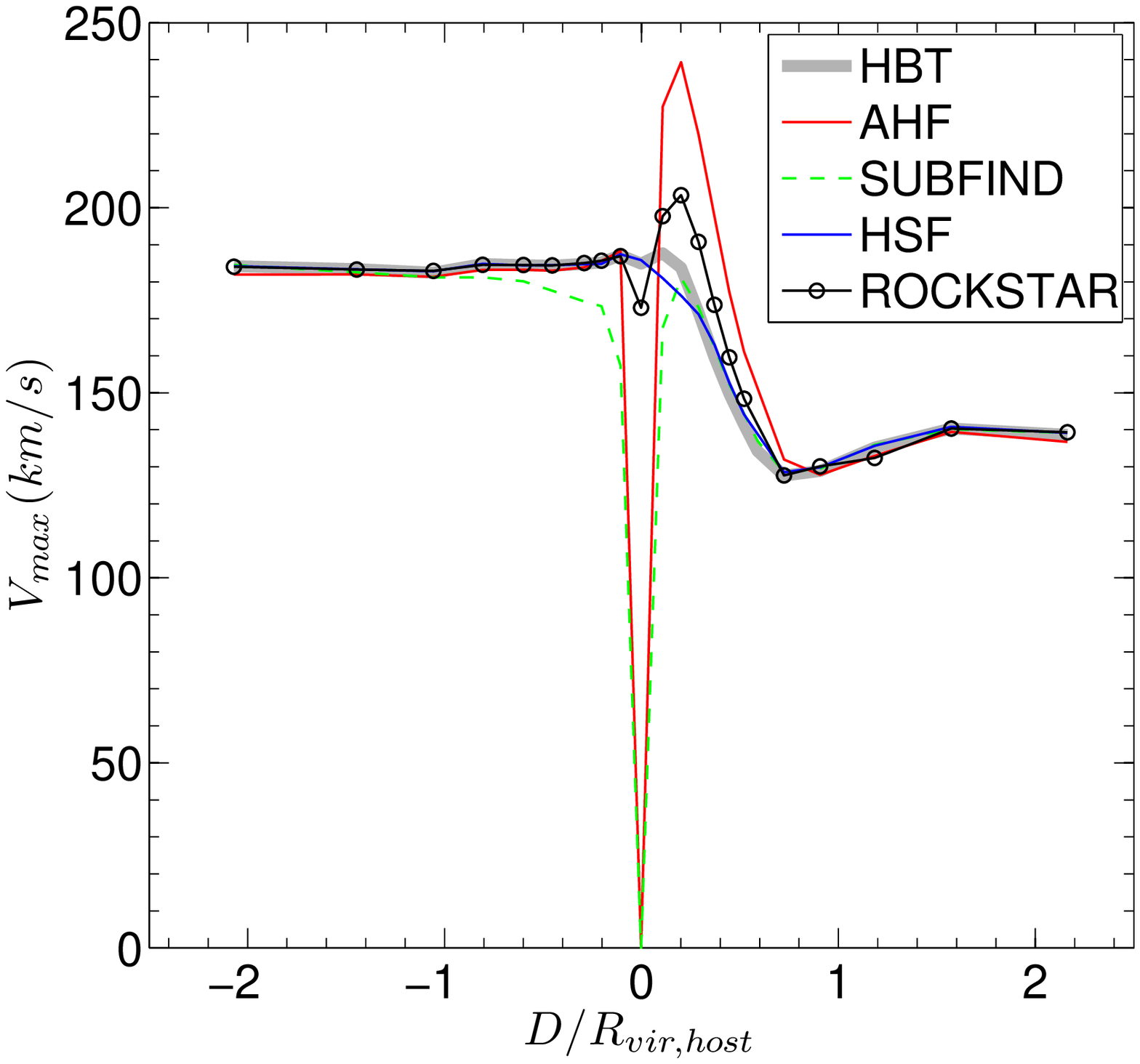}
\caption{Mass stripping history of a subhalo falling through the
  centre of a host halo as found by HBT and the other four representative
  subhalo finders. Left: Bound mass of the subhalo normalized by its
  virial mass before infall, as a function of its distance from host
  centre, normalized by the virial radius of the host halo. The
  subhalo moves toward the positive direction of the axis. Right:
  Evolution of the maximum circular velocity of the
  subhalo. \rev{The circles on the ROCKSTAR line mark the timesteps of the simulation outputs.}}\label{f:mad}
\end{figure*}

\subsection{Subhalo Mass Stripping History}\label{s:MAD}
The test simulation we use from the "haloes Gone Mad" project 
is a head-on collision simulation of two NFW haloes with initial virial
masses of $10^{12} \rm{M}_\odot$ and $10^{14} \rm{M}_\odot$,
corresponding to $10^4$ and $10^6$ particles. The small halo is thrown
right through the centre of the other halo. As shown in \citet{Mad},
all halo finders based on only configuration space information fail to
find the subhalo at host centre, and even some finders based on
position and velocity phase space information give an un-physical
result near the centre. For clarity, we choose two representative
configuration based finders (AHF and SUBFIND) and two phase-space
based finders (HSF and ROCKSTAR), and compare our HBT result in
Figure~\ref{f:mad} with those found by the four finders. It can be
seen that the HBT subhalo is complete from the start, robust near the
halo centre, and clean as it moves out of the central region. As just
stated, the two configuration based finders fail to find the subhalo
at the very central region of the host halo. HSF gives an unphysically
high subhalo mass in the beginning (higher than the halo mass before
the infall), which may be attributed to the inclusion of some local
particles as discussed in section~\ref{s:acc}. AHF gives a much higher
mass when the subhalo just passes the centre, which is not physical
because, as can be seen from the right panel, the maximum circular
velocity $V_{max}$ well exceeds that of the progenitor halo. A
significant fluctuation in the $V_{max}$ history is also observed for
the subhalo found by ROCKSTAR near the host centre, indicating some
ambiguity in capturing the bound mass by the finder. Furthermore,
there is no obvious reason why the subhalo's mass should peak at 0.4
times the host virial radius as given by ROCKSTAR. In general HBT has
comparable performance with HSF, while still has the advantage that
all the HBT particles are strictly physically associated with the
subhalo by design. 

\subsection{Subhalo mass function}\label{ss:msfun}
\begin{figure}
\myplot{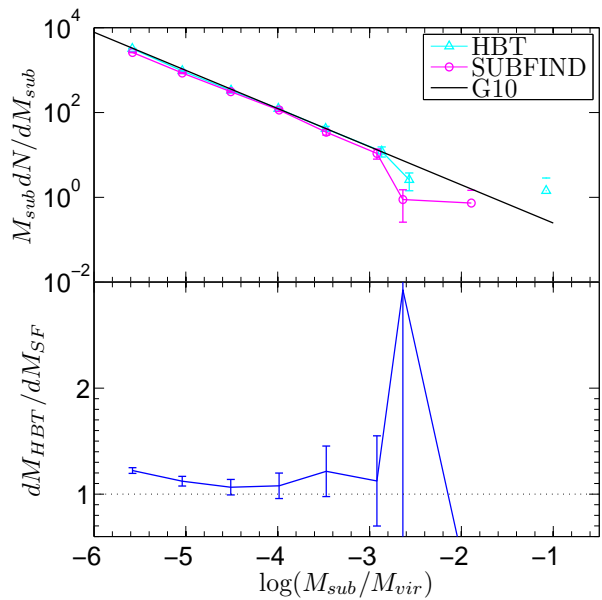}
\caption{Subhalo mass function at z=0 for the cluster resimulation. Top
panel
  shows mass weighted subhalo mass function as found by HBT and
  SUBFIND. The black solid line is the fitting formula of 
  Equation~\ref{eq:mfun} with the normalization given by
  \citet{G10}. Bottom panel plots the ratio of the total subhalo mass contained
  in each mass bin from the two codes, or equivalently the ratio of
  the mass-weighted subhalo mass functions. The horizontal dashed line
  is the 1:1 reference.}\label{f:mfun_resim}
\end{figure}

In Figure~\ref{f:mfun_resim} we show the subhalo mass function found
by HBT and SUBFIND for the resimulated cluster at z=0. For reference
we also plot the power-law fitting formula
\citep{Gao04b,Aqua,Angulo,G10}.
\begin{equation}\label{eq:mfun}
\frac{dN}{M_{host} d\ln(M_{sub})}=N_0 M_{sub}^{-0.9}
\end{equation}
where $M_{sub}$ is subhalo mass and $M_{host}$ is the host halo virial
mass. The normalization $N_0$ depends on the definition
of the virial radius within which subhaloes are counted. For the virial
relation predicted by spherical collapse model\citep[see e.g.][for a
fitting formula]{Bryan}, G10 found the
normalization to be $N_0=10^{-3.03}(h^{-1}\rm{M}_\odot)^{-0.1}$. We
adopt the same virial definition throughout this paper, \rev{with the average density within the virial radius defined to be 101 times the critical density at $z=0$,} and will
compare our subhalo mass function with the G10 result. It can be seen
that the mass functions are consistent with the fitting formula from
G10, with HBT's subhaloes being more massive than the result of SUBFIND by
10 to 20 percent.

\subsection{Size of Subhaloes}\label{ss:size}
\begin{figure}
\includegraphics[width=0.5\textwidth]{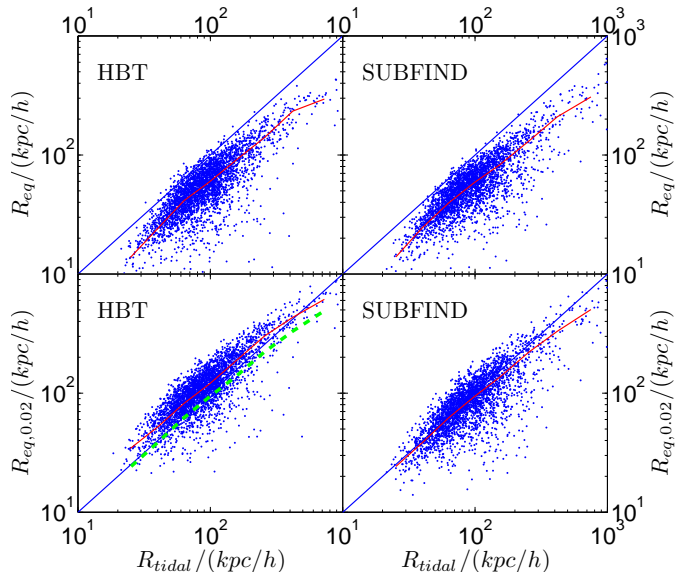}
\caption{Size of subhaloes in the 100 most massive haloes in the
  cosmological simulation as found by HBT and SUBFIND. $R_{eq}$ and
  $R_{eq,0.02}$ are the radii where the subhalo density equals to 1 and
  0.02 times the background density. $R_{tidal}$ is the subhalo tidal
  radius. We use only FoF haloes containing more than 1000 particles
  and subhaloes containing more than 100 particles. The red solid line
  in each panel marks the median value of $R_{eq}$ (or $R_{eq,0.02}$
  according to y-axis). In the lower left panel we overplot the median
  line of $R_{eq,0.02}$ of SUBFIND result in green dashed
  line.}\label{f:8213size}
\end{figure}
The size of a subhalo is trimmed by the tidal force from the host halo,
and can be estimated by a tidal radius which is defined as the radius
of the satellite subhalo at which its self-gravity equals the tidal
force of the host halo. In the limit $R_{tidal}<<D$ where $D$ is the
halo-centric distance of the subhalo, the tidal radius is estimated
as\citep{GalacticDynamics,Tormen98}:
\begin{equation}
R_{tidal}=D\times\left[\frac{M_{sub}}{(2-\frac{d\ln M_{host(<D)}}{d\ln
D})M_{host(<D)}}\right]^{1/3} \label{eq:tidal}
\end{equation}

For singular isothermal density profile, it is easy to check from
Equation~\eqref{eq:tidal} that the local density of the host halo
equals to that of the satellite at tidal radius. Thus a radius of
equality $R_{eq}$ can be defined at which the density of the satellite
is the same as the local density of the host. For realistic density
profiles, the tidal radius may differ from $R_{eq}$ by a factor of
order unity. In \citet{Aqua} they found that the subhalo tidal radius
equals to a radius $R_{eq,0.02}$ where the subhalo's density falls
below 0.02 times the local host density. We compare the tidal radius
to $R_{eq}$ and $R_{eq,0.02}$ for both HBT and SUBFIND results in
Figure~\ref{f:8213size}, using the 100 most massive FoF haloes in the cosmological simulation. To avoid resolution effect, we use only FoF
groups with more than 1000 particles and subhaloes with more than 100
particles. It can be seen that although the $R_{eq,0.02}$ is very
close to $R_{tidal}$ for SUBFIND result, it is in general $20\sim30$
percent larger in the HBT case. But for $R_{eq}$ the two codes give
consistent results. This reflects a puff-up in the outer region of
subhaloes in HBT result, or rather a sharp cut-off in the SUBFIND
result, due to particle division using density saddle points. We show
this explicitly in the next subsection.

\subsection{Density Profile}\label{ss:profile}
\begin{figure}
\includegraphics[width=0.5\textwidth]{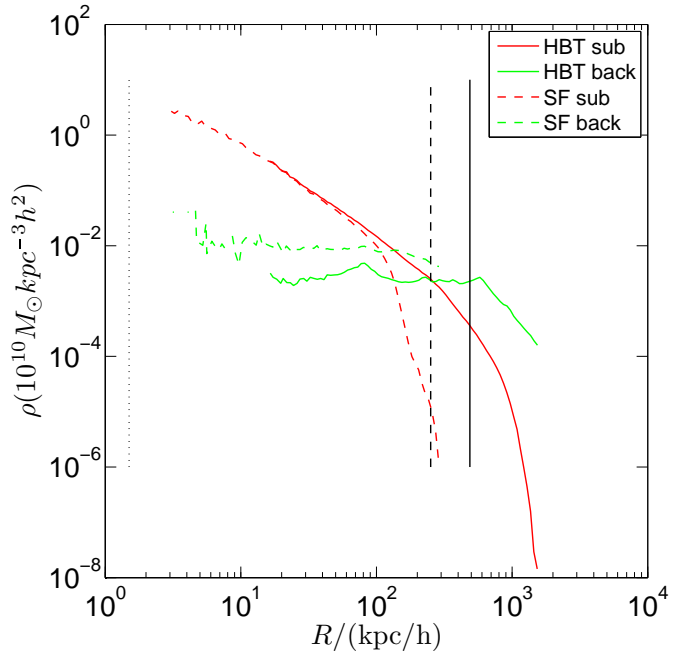}
\caption{Density profiles for the most massive satellite subhalo of the
  resimulated cluster as found by HBT and SUBFIND. Red lines are for
  particles which are bound to the subhalo, green lines are the total
  density of all the other particles. The vertical black lines mark
  the tidal radius of the subhalo. Solid lines are the HBT result
  while dashed ones are the SUBFIND result. The vertical dotted line
  on the left marks the smoothing length of the
  simulation.}\label{f:6702satprof1}
\end{figure}

There is almost no difference in the density profile between central
subhaloes found by HBT and SUBFIND. But for satellite subhaloes,
especially when they reside in the central region of host haloes, HBT
gives a much more extended profile. Figure~\ref{f:6702satprof1}
compares the density profiles of the most massive satellite of the
cluster found with the two codes. This subhalo is resolved to have
$10^6$ particles or $1/10$ the virial mass of the cluster. It is
located at $0.3R_{vir}$ from the cluster centre. As expected, a sharp
cut is observed for the SUBFIND profile near its tidal radius, while
HBT gives a subhalo mass which is 6.5 times as large. This is due to
HBT's prior knowledge of source particles beyond the tidal radius
while SUBFIND can only identify those particles within tidal radius,
as already seen in section~\ref{ss:size}.

\subsection{Cross Match between HBT and SUBFIND}\label{sec:match}
To see if the two codes find the same set of subhaloes, we try to match
the members in the two catalogues. Given one target subhalo in hand,
we search host subhaloes for its member particles in the other
catalogue. For each host subhalo, we calculate a boundness weighted
summation of the number of matched particles, with the most-bound
particle in the target subhalo having the largest weight.
The host subhalo with the biggest summation is selected as the
target's correspondence in the other catalogue. We do the match from
HBT to SUBFIND and vice versa. We classify the subhaloes into three
categories according to the two matches: bilaterally matched subhalo
when the subhalo and its correspondence are matched to each other;
unilateral subhalo when the correspondence cannot be matched back to the
subhalo; and un-matched subhalo when no correspondence can be
found in the other catalogue.

\begin{figure}
\includegraphics[width=0.5\textwidth]{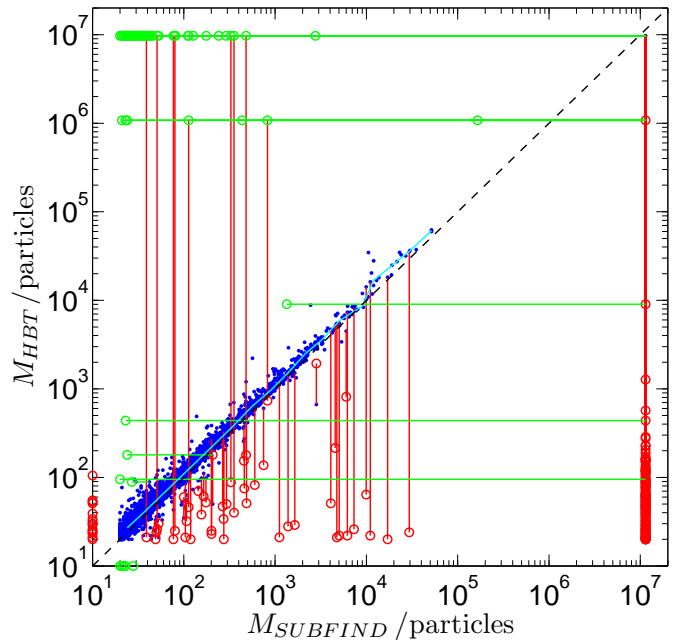}
\caption{Cross match between HBT and SUBFIND subhaloes for the
  cluster. Each subhalo is plotted according to its mass and the mass
  of its correspondence. Blue dots are bilaterally matched subhaloes
  (see text for definition). The solid cyan line gives the median masses
  of the bilateral subhaloes.  Red circles are unilateral HBT subhaloes
  and green squares are unilateral SUBFIND subhaloes, except for those
  on the axes which are unmatched ones. For each unilateral match, we
  also draw a line connecting that subhalo to its
  correspondence. The point at the top-right corner is the central subhalo.}\label{f:6702match}
\end{figure}

The match result for the galaxy cluster is shown in
Figure~\ref{f:6702match}. The majority of subhaloes are bilaterally
matched, accounting for 82\% HBT subhaloes and 97\% SUBFIND
subhaloes. 90\% of the bilateral SUBFIND subhaloes have more than 95\%
of their particles shared with their HBT correspondences. The median
line points out 10 to 20 percent increase in mass for HBT subhaloes
compared to their SUBFIND correspondences. Only less than 1 percent
subhaloes belong to the un-matched category, with their masses being lower
than 100 particles. 

\begin{figure}
\includegraphics[width=0.5\textwidth]{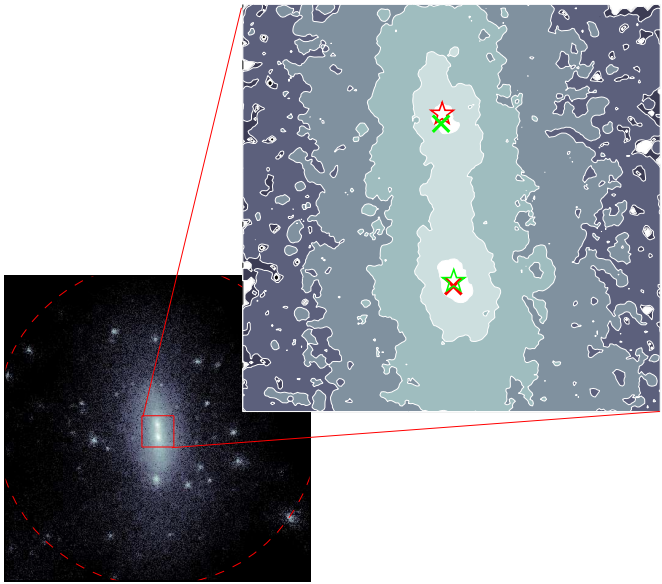}
\caption{An example of sub-in-sub switch. These are two subhaloes
  forming a binary system near the boundary of the host halo. The lower-left
   panel shows the projected squared density of this system. The 
   red dashed circle in the lower-left panel marks the virial radius of the system.
   The upper-right panel shows the contour map of the density field for the very inner region. 
   Red symbols mark the most-bound particles of two HBT subhaloes; green symbols mark those
   of SUBFIND subhaloes. Pentagrams represent the superior subhaloes and
   crosses stand for subordinate subhaloes. The masses for the upper and
   lower subhaloes in the image are 94468(488) and 17541(111884)
   particles in the HBT(SUBFIND) case.}\label{f:switch}
\end{figure}

Unilateral subhaloes are discussed extensively in Appendix~\ref{sec_uni}.  Here
we only focus on one particularly interesting case. When two haloes with
comparable masses merge, it is likely that the resulting halo contains two
subhaloes with comparable peak densities, making it difficult to distinguish
which is superior and which is subordinate without appealing to progenitor
information. Especially when these two subhaloes are close to each other forming
a high density environment, the difficulty of resolving and weighing them are
elevated. We give an example of such a binary system in Figure~\ref{f:switch}
from the cosmological simulation. Two satellite subhaloes sit at the centre of
the image resulting from a
 1:2 merger of two progenitor haloes. This binary satellite system is found near
   the boundary of the host halo and thus closely resembles an individual halo.
   Most of the surrounding particles are bound to both of the density peaks.
   SUBFIND takes the peak from the smaller halo which has higher central
   density and associates the surrounding particles with it to make a parent
   subhalo, leaving the core from the bigger halo as its sub-in-sub, while HBT
   gives a reversed sub-in-sub hierarchy which is consistent with the merger
   hierarchy. When we match these two subhaloes from SUBFIND to HBT, obviously
   the small subhalo in SUBFIND is matched to the big one in HBT. And because
   the surrounding mass outweighs the core in the big subhalo of SUBFIND, this
   one is also matched to the big subhalo in HBT. For the same reason, the two
   HBT subhaloes are also matched to the big SUBFIND subhalo. This may be of
   particular importance when constructing merger histories. A false switch
   between two branches of a tree would happen as a result of the switch
   between the sub-in-sub relation of the binary. With the help of subhaloes'
   merger history, HBT is able to give more physical weights to the dominance
   of the two cores and avoid this kind of faulty links in the merger tree. In
   Figure~\ref{f:6702match}, these binary subhaloes would produce two adjacent
   perpendicular lines connecting near the 1:1 line if the surrounding
   particles could outweigh one core, or a single line if one subhalo is
   missing. A dedicated study of how major-mergers challenge different subhalo
   finders can be found in Behroozi et al.(in prep).
  
\section{Summary}\label{s:conc}
In this work, we have managed to develop a tracing algorithm to
produce a complete and physically-motivated subhalo catalogue. In our
HBT algorithm, subhaloes are divided into two types where central
subhaloes grow via accretion from the host halo and from satellites
inside the same host while satellite subhaloes can only accrete from
within their satellite-of-satellites. We keep a full record of subhaloes'
merger hierarchy and apply the unbinding algorithm hierarchically to
enable satellite accretion and merger. While omitting satellite
accretion can result in 20 percent loss in mass for massive
satellites, individual tests of subhalo mass loss history and
statistical comparison with SUBFIND which is based on local density
search show that local accretion of background particles have only
temporary effect and is negligible for satellite subhaloes.

Because of the large dynamic range in the evolution of subhaloes' masses,
a robust unbinding algorithm is required to trace subhaloes long
enough. We have proposed core-averaged unbinding algorithm together
with self-adaptive update of source subhaloes to accomplish this
job. The core-averaged unbinding algorithm succeeds in its ability to
give quick and accurate estimation of the centre and bulk velocity of
the subhalo, allowing safe removal of unbound particles. Self-adaptive
update of source subhaloes keeps the dynamic range in a safe region for
unbinding while allowing rebinding of stripped particles. We find
that successive tracing of satellites without allowing rebinding can
cause up to a factor of 3 suppression in the subhalo mass at apocentre
passage. Application of HBT to a contrived simulation also
demonstrates its superb performance in robustly recovering the
subhalo's stripping history.


Since our algorithm utilizes historically constructed source subhaloes rather than
locally collected sources using density and position thresholds, we
can resolve satellite subhaloes well even in the central high density
region of haloes while SUBFIND may have severe spatial truncation for
subhaloes in halo centre. Even for small satellites our subhalo mass
function is about 15 percent higher than SUBFIND. The size of subhaloes
found by HBT extends 20 to 30 percent larger than SUBFIND
characterized by $Req_{0.02}$. 

Because we do not need to do density interpolation or spatial searching to
construct source subhaloes, HBT runs fast. The HBT code is written in C and has
been fully OpenMP parallelized both for cosmological simulations where the
parallelization is on halo level and for high resolution resimulations where
the parallelization is done on particle level; a hybrid MPI/OpenMP version is
under development and will be available in the near future. It would also be
straight forward and efficient to integrate HBT to a simulation code for on-the-fly
 high resolution subhalo finding and merger tree output. In that case HBT
would also not be limited by the number of snapshots in the simulation output.

\section{Discussion}\label{s:discussion}
 Much of the treatment in HBT is more physical than mathematical. Its
 complexity lies in the physical process involved and its success
 reflects its correct understanding of the subhalo's evolution. This code
 is obviously not applicable to hot dark matter simulations, where
 structure formation is top-down rather than bottom-up. We would also
 expect HBT to have some difficulty when applied to warm dark matter
 simulations, where fragmentation is a vital process in forming
 structures\citep{warm}, though our splitting algorithm would
 alleviate the problem. In fact HBT can be regarded as a low-level
 semi-analytic model for cold dark matter subhaloes. Its high
 resolution and physical particle partitioning guarantees that the
 merger trees built by HBT have much fewer lost nodes or false
 links. This makes HBT an ideal choice for building merger trees for
 Semi-Analytic Models of galaxy formation. \rev{Besides, the tracking 
 nature makes HBT easily extensible to find not only bound subhalos, but
 also more general structures such as streams. A combination of HBT's tracking
 ability with the STructure Finder's \citep[STF;][]{STF} stream identification algorithm
 is under development to trace streams (Elahi et.al. 2012, in prep).}

One example revealing HBT's advantage for galaxy formation models is the
"sub-in-sub switch" case shown in Figure~\ref{f:switch} and discussed in
section~\ref{sec:match}.  When two subhaloes of comparable mass form close
pairs, HBT is superior in its ability to partition the surrounding particles
according to their origins. This is important because the surrounding
particles, while being co-bound by the binary-subhalo and can be assigned
arbitrarily to either member, play a vital role in establishing links to
subhaloes' progenitors and descendants, especially when they outweigh the cores
of the binary. HBT's way of partition would guarantee that these surrounding
mass obey the progenitor-descendant relation, while a partition without knowing
the origin of these particles can easily lead to an incorrect link of the cores
to the progenitor haloes. 

 One future direction for the improvement of HBT code, and other
 subhalo finders as well, would be to find a better definition for
 subhaloes. Even with our ability to clearly construct source subhaloes
 near halo centre, subhaloes can still appear over-stripped near
 pericentre after which some stripped mass get rebound, posing a
 remaining "resolution" problem. We argue this remaining ``resolution"
 problem to be due to the operational definition of subhaloes as
 instantaneously self-bound structures, which is currently adopted by
 almost every subhalo finder. This definition only guarantees
 coherence for a structure in isolation and in the collisionless limit
 of its member particles. Putting aside the structural evolution of
 the subhalo itself, there is still an evolving background which
 lowers the gravitational potential of the subhalo while at the same
 time produces tidal force. In \citet{Shaw07} both the tidal force
 from the host and the gravitational potential of stripped particles
 are considered to improve the coherency of subhalo definition. They
 find that the contribution from the tidal force and that from the
 lowered potential due to background particles roughly offset each
 other, with a slight increase for the mass of subhaloes in the inner
 region of a halo.

It should be noted that tidal force is not always disruptive. This
could easily be illustrated in the following simple picture. Consider
two test particles with infinitesimal mass moving on the same
elliptical orbit around a central mass $M$, starting from apocentre
with a small time delay $dt$ between them. The relative velocity is
then $d\overrightarrow{v}=\frac{GM}{r^3}\overrightarrow{r}dt$. Hence
the internal kinetic energy of the system varies as $dv^2\propto
r^{-4}$, reaching a minimum at apocentre and maximum at
pericentre. Because the internal gravity of the two test particles can
be ignored, the change in the internal kinetic energy is solely from
work done by tidal force. It is ready to see that the tidal work helps
to reduce the relative kinetic energy of the two particles from
pericentre to apocentre. As a result, particles previously marked as
unbound to a subhalo can become bound again as the subhalo moves to
the outer region of its host. This has been seen in our case study of
the evolution of individual subhaloes in section~\ref{s:robust}, where
strong rebinding is observed as satellites move out from the centres
of haloes.

We comment that in \citet{Shaw07} even though tidal energy has been
included in the unbinding procedure, they remove velocity outliers
before adding tidal energy, still failing to allow these high velocity
particles to be re-captured through tidal work.

\section*{Acknowledgements} 

We thank Volker Springel for making the SUBFIND code available, and the
anonymous referees for lots of helpful comments. This
work in Shanghai is supported by NSFC (10821302, 10878001, 11033006),
by 973 Program (No.2007CB815402), and by the CAS/SAFEA International
Partnership Program for Creative Research Teams (KJCX2-YW-T23). HYW is
Supported by NSFC 11073017. During the final stage of this work, JXH
is supported by the European Commission’s Framework Programme
7,through the Marie Curie Initial Training Network Cosmo-Comp
(PITN-GA-2009-238356).

\clearpage
\appendix
\section{Unbinding Algorithm}\label{sec_unbind} 
The core-averaged unbinding in HBT is done in the following steps.

\begin{enumerate} 

\item
Starting from an initial assemble of $N$ particles of a source subhalo.  

\item
We calculate the gravitational potential $\psi$ for each particle from the
contribution of all the other $N-1$ particles. We use the Barnes-Hut tree
algorithm \citep{BH1,BH2} as implemented in GADGET\citep{GADGET,GADGET2} to
calculate the potential. We then find a fraction (set by a parameter
\texttt{CoreFrac}) of particles with the lowest potential, and call it as the
lowest potential core.  

\item We calculate the kinetic energy $K$ of each
particle, including Hubble flow, with respect to the average velocity and
centre of mass of the lowest potential core. Then we remove any particles with a
positive total energy $K+\psi>0$.  

\item If the remaining number of particles
$N_b$ is below the user desired mass limit \texttt{NBoundMin}, the unbinding
procedure stops with no subhalo found.  Otherwise, if $N_b$ agrees with $N$
within some accuracy \texttt{MassPrecision}, unbinding stops with these $N_b$
particles as a subhalo. In the other case, take the remaining $N_b$ particles
as an initial assemble of particles and repeat the above steps 2-4.

\end{enumerate} 

The parameter \texttt{MassPrecision} is introduced to improve the efficiency of unbinding, which is the most time-consuming part of the entire code. Strictly speaking the definition of self-boundness
requires the iterative removal of un-bound particles to be 100 percent
complete, i.e, the iteration can only stop when the number of bound
particles $N_{bound}$ equals to the number of remaining particles
$N$. However, as we have discussed in section~\ref{s:discussion},
self-boundness is not a perfect definition to identify particles
associated with a subhalo. Practically, it would be enough to just
give an estimation of the bound structure which is fairly close to the
final self-bound part.

\begin{figure}
\includegraphics[width=8cm,height=8cm]{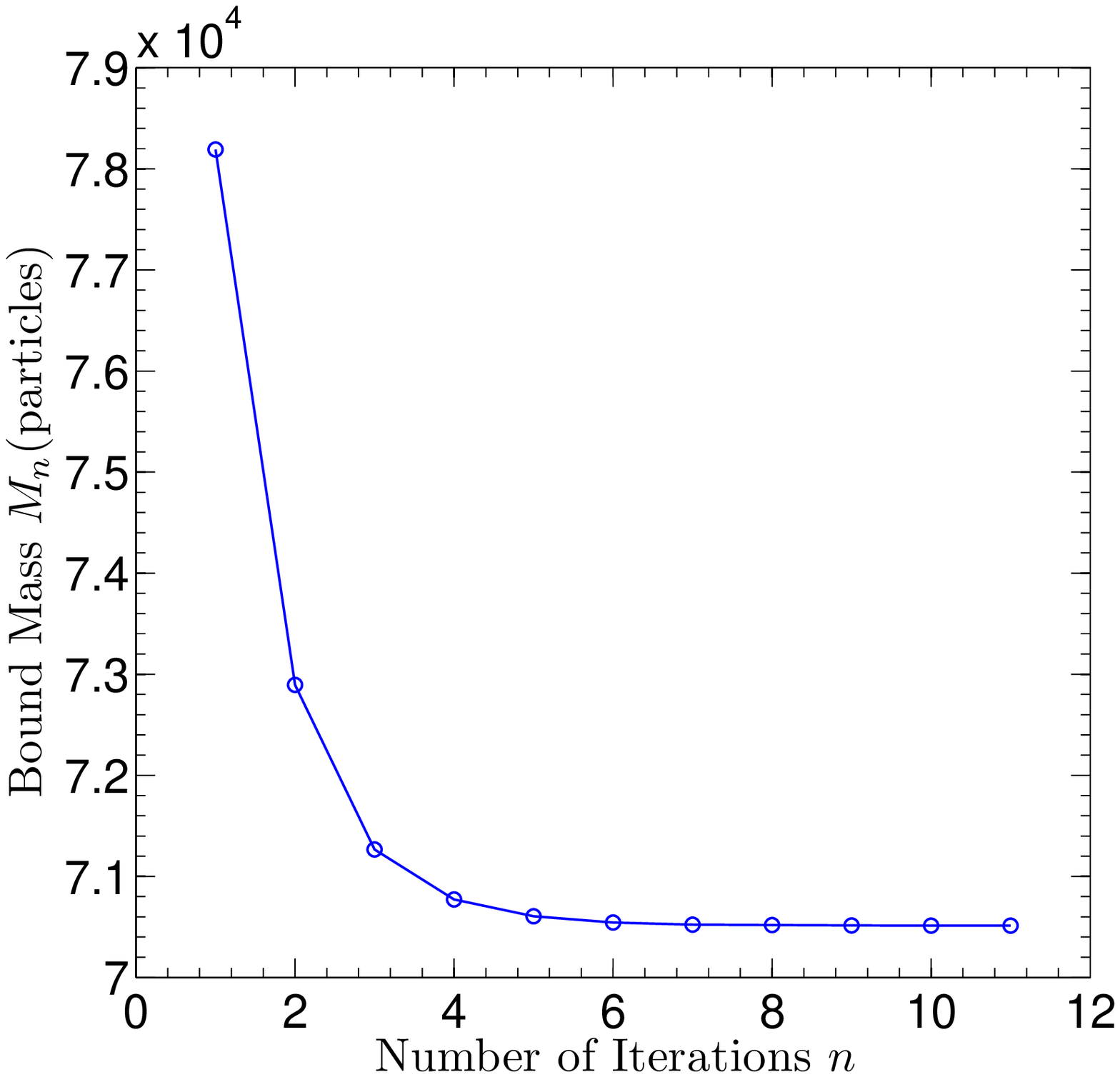}\\
\includegraphics[width=8cm,height=7.7cm]{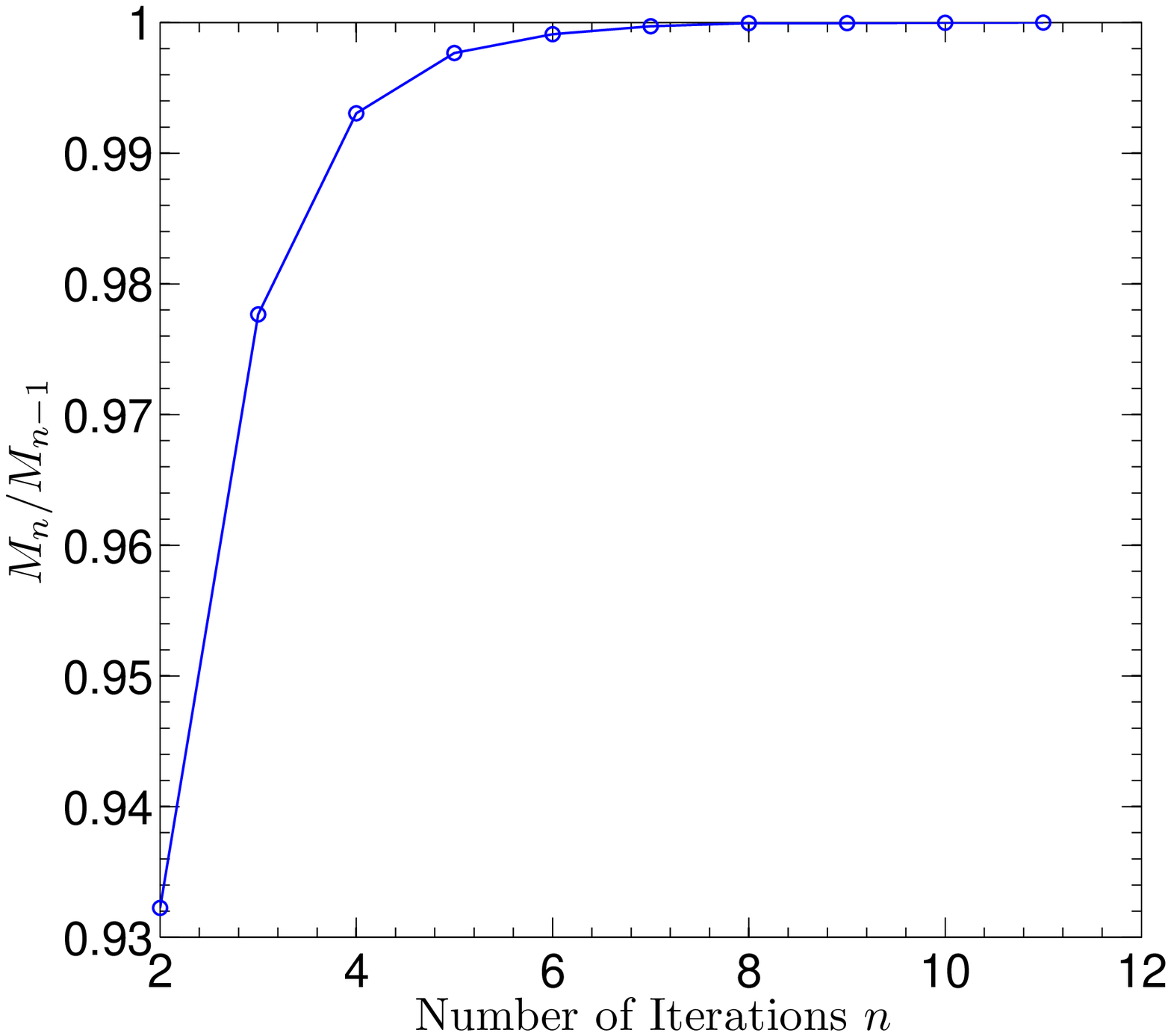}
\caption{Typical convergence curve of a subhalo during unbinding. Top: The
remaining bound mass after each iteration; Bottom: Estimated precision of
the bound mass after each iteration.}\label{f:converge}
\end{figure}

We investigated the speed of convergence for several randomly selected
subhaloes and find that the bound mass converges quickly with the
number of iterations towards the final self-bound mass, as shown in
Figure~\ref{f:converge}. After several iterations, the majority of
unbound particles have already been removed. Further iterations only
refine the bound mass to a higher accuracy, with smaller and smaller
change in mass as the iteration goes on. An estimation of the
precision of bound mass can be obtained as the ratio of bound mass at
two subsequent iterations. By stopping the iteration at some precision
threshold \texttt{MassPrecision}, a lot of iteration steps can be
saved while keeping the mass estimation accurate to the pre-set
accuracy. \rev{We adopt $\texttt{MassPrecision}=0.995$ in the current 
implementation of HBT.}

\section{Possible Local Accretion}\label{s:acc}
\begin{figure}
\myplot{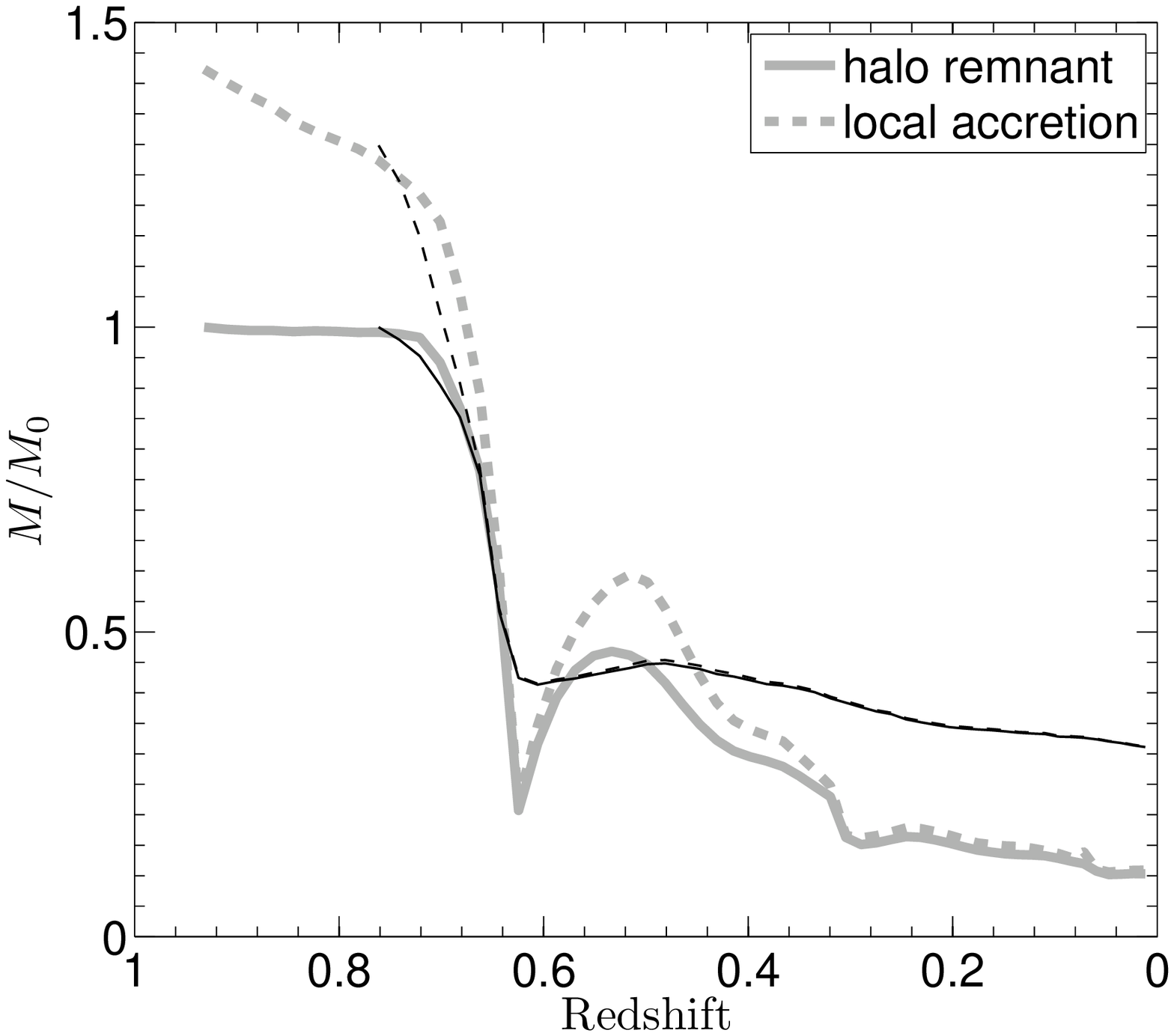}
\caption{Effect of local background particles vs. infalled source
  subhalo. The solid line shows the bound mass from the infalled halo,
  as done in the right panel of Figure~\ref{f:robustness} with
  $\mathtt{CoreFrac0}=0.25$. The dashed line shows the mass after
  adding bound local particles. The masses are normalized by the
  subhalo's mass at infall. Thick grey and thin black lines are for
  haloes S43G11 and S51G86 respectively. }\label{f:local_acc}
\end{figure}

The physical assumption under tracing algorithms is that the mass
accretion
of satellite subhaloes can be ignored within a host halo, so that one
needs only to follow the remnants of infalled haloes. In
Figure~\ref{f:local_acc} we give a direct test of this
assumption. After extracting a self-bound subhalo from its adaptive
source, we search for and add any additional bound particles located
within $2$ times its virial radius (defined as if the subhalo is a
halo). Except in the first several snapshots (and the first apocentre
passage in the S43G11 case) when the subhalo is around the boundary of
its host, no accretion of local background mass is found.

\section{Maximizing the Resolution of haloes and Subhaloes}
\subsection{Splitting algorithm}\label{ss:split}
There are several conditions under which one may need to split a source
subhalo into parts. One case is that a source subhalo may contain
additional substructures which are unresolved given the time and
spatial resolution of the simulation outputs, or from a casual link of
two haloes as one. These unresolved substructures may separate from the
hosting source subhalo later in orbits, due to orbital parameter
difference or tidal forces of nearby structure. Also a strong tidal
force can break up a nearby smooth halo. In these cases a source
subhalo may have multiple descendants hosted by several haloes. We
implement the following splitting algorithm to take this into account:

For each source subhalo, we partition its particles into fragments
according to their current host haloes. Fragments with particle numbers
smaller than \texttt{NSrcMin} are discarded. Here \texttt{NSrcMin} is
a parameter controlling the mass resolution of source subhaloes. We
choose this to be equal to \texttt{NBoundMin} in our
implementation. If all the fragments are discarded, discard this
source subhalo. The subhalo out of the most massive fragment is taken
as the ``main descendant'' of this source subhalo. Other remaining
fragments are called splinter source subhaloes, the subhaloes directly
out of which are named splinter subhaloes. We keep a record of both the
main descendant and the splinter information, yielding a merger tree
which can have ``downward branches''. Note that because we do not
appeal to spatial clustering within haloes to identify subhaloes, our
tracing algorithm can only resolve splitting when it happens spanning
over several haloes. Within one halo even if a subhalo breaks into two
obvious part, as long as they are still bound as a whole we have no
way to split them in HBT.

The mass function of the biggest splinters (excluding the main
descendant) in each splitting event decays approximately as
$\frac{dN}{d\ln m}\propto m^{-2.5}$, much steeper than the slope of
subhalo mass function $\frac{dN}{d\ln m}\propto m^{-0.9}$. It's ready
to see that these splinters are primarily small subhaloes near the
resolution limit.

\subsection{Quasi-halo treatment}
We take those subhaloes for which we cannot find their host FoF haloes as being in
a ``background halo'' when assigning hosts in tree constructing
algorithm as well as when splitting source subhaloes. This way we can
find subhaloes which do not have a known host halo. Their source
subhaloes can be regarded as ``quasi-haloes'' which are not identified
by the halo-finder, but they still have concentrated structure and
progenitor-descendant link between neighboring snapshots. These
quasi-haloes are haloes fluctuating around the resolution limit. The
biggest quasi-haloes identified have about 50 particles when haloes are
filtered by a mass limit of 10 particles in our
implementation. Although they could be split or ejected subhaloes,
those who can grow to a meaningful size are almost all haloes at their
earliest stage of growth, i.e, haloes which have just become
resolvable. The introduction of quasi-haloes is intended to recover
the lost nodes in the merger tree due to halo-finder pitfalls, but
has little effect on the tree even if omitted because they come as
primarily early fluctuations.

\subsection{Time resolution dependence}\label{s:time_resolution}
In our tracing algorithm, one problem of concern is whether the time
resolution of simulation outputs would affect the tracing result. The
effect of time resolution is two-fold: time resolution translates
into resolution of halo growth history; it also changes the dynamic range 
for unbinding thus affecting its robustness.

 First, there could be un-resolved satellites as well as un-resolved 
 growth of satellites. Obviously those haloes born between two subsequent
 outputs and immediately becoming satellite subhaloes in the second snapshot are
not recorded in halo catalogues and their descendants as independent
subhaloes are missed. However, since these haloes get stripped right
after they become discernible, they do not have time to grow and are
always fairly small haloes. Also the better the time resolution and the
smaller the halo mass limit, the smaller are these unresolved
haloes. For those haloes which merge between two subsequent
snapshots, the growth of satellite haloes after the first snapshot and
before merger are not captured either. Thus one would expect a
decrease in subhalo population with decreasing time resolution.

\begin{figure}
\includegraphics[width=0.5\textwidth]{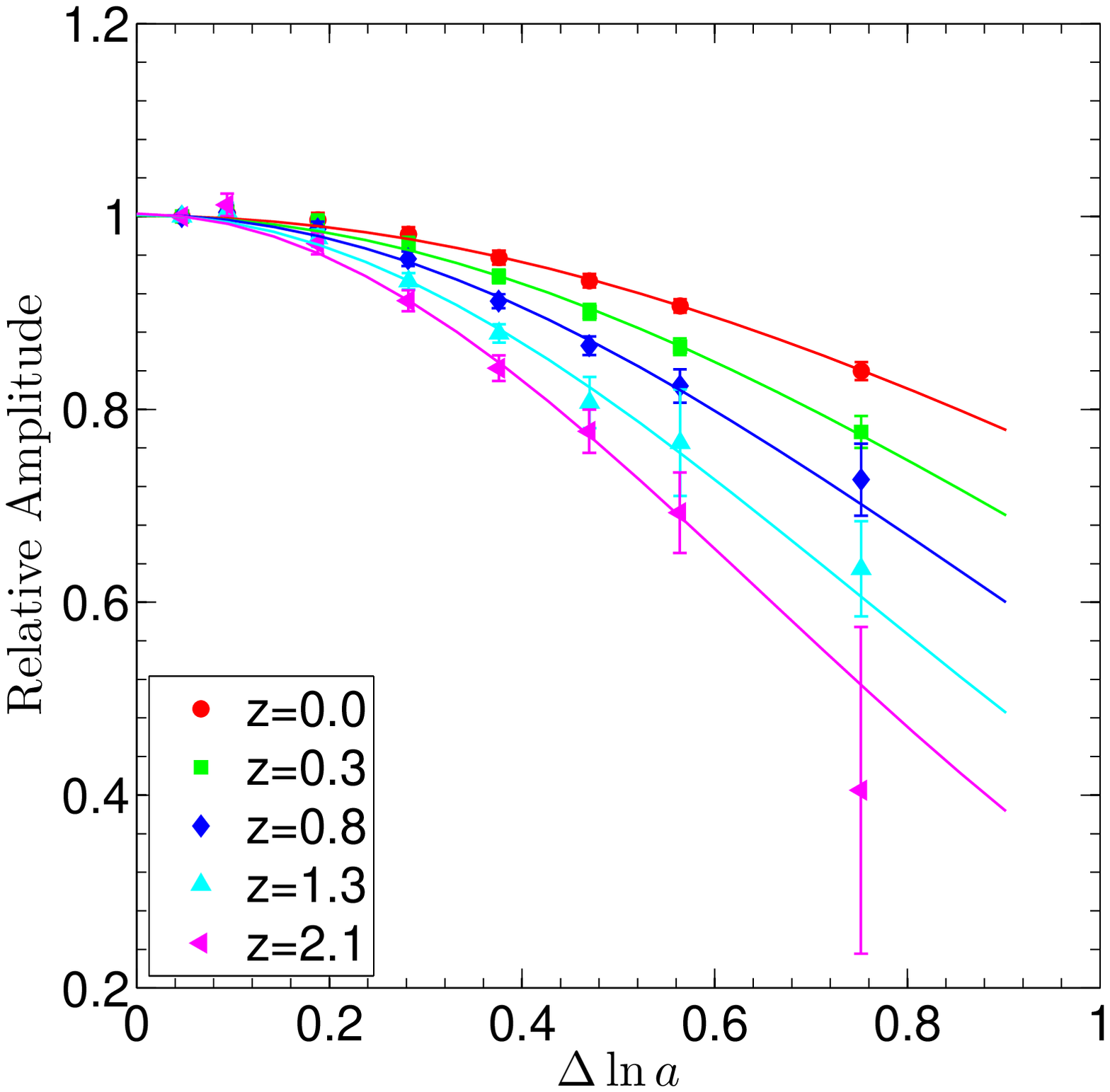}
\caption{The Effect of time resolution on subhalo abundance. Data
  points are the relative amplitudes of the diluted subhalo mass
  functions to the best resolved one, as a function of time
  resolution. We show the result at five different redshifts, together
  with a Gaussian function fitting at each redshift. Error bars in the
  figure are propagated from Poisson errors on the subhalo mass
  functions. }\label{f:mfun_time_res}
\end{figure}

 To quantify this effect, we compare the subhalo mass function
 $M_{host}^{-1}dN/d\ln M_{sub}$ identified with different time
 resolution. Our cosmological simulation has 60 outputs from $z=15.02$
 to $z=0$ equally spaced in log-space of the scale factor. We dilute
 these outputs by keeping one snapshot after skipping every a few
 number of snapshots. Subhaloes are identified using these diluted
 snapshots with HBT. It is found that the diluted subhalo mass
 function has the same shape as the best resolved one, with the ratio
 of them well fitted by a constant and depending weakly on host halo
 mass, as long as the time step used for tracing is not too large. In
 Figure~\ref{f:mfun_time_res} we examine the relative amplitude of the
 diluted subhalo mass function to the best resolved one (with
 $\Delta\ln a=0.0468$) at five different redshift for the cosmological
 simulation. The relative amplitude $A$ as a function of time
 resolution can be well fitted by a Gaussian function
\begin{equation}\label{eq:mfun_time_res}
A=A_0e^{-(\frac{\Delta\ln a}{b})^2}
\end{equation}
with parameters $A_0$ and $b$, where $\Delta\ln a$ is the difference
in $\ln a$ between subsequent snapshots with $a$ being the scale
factor. The parameter $A_0$ represents the relative amplitudes
extrapolated to infinitesimal time resolution. The fitted $A_0$ are
all fairly close to unity, increasing from $1.001\pm0.001$ to
$1.003\pm0.001$ from $z=0$ to $z=2.1$, indicating that the best
resolved mass function is over $99.7\sim 99.9$ percent complete. The 
parameter $b$ depends on the scale factor as $b=1.29a+0.51$. Equation~\ref{eq:mfun_time_res}
provides a handy reference in assessing the completeness $A/A_0$ for
subhalo populations from HBT, although the exact completeness for a particular
application would also depend on the mass resolution of the halo and subhalo catalogues.

\begin{figure}
\includegraphics[width=0.5\textwidth]{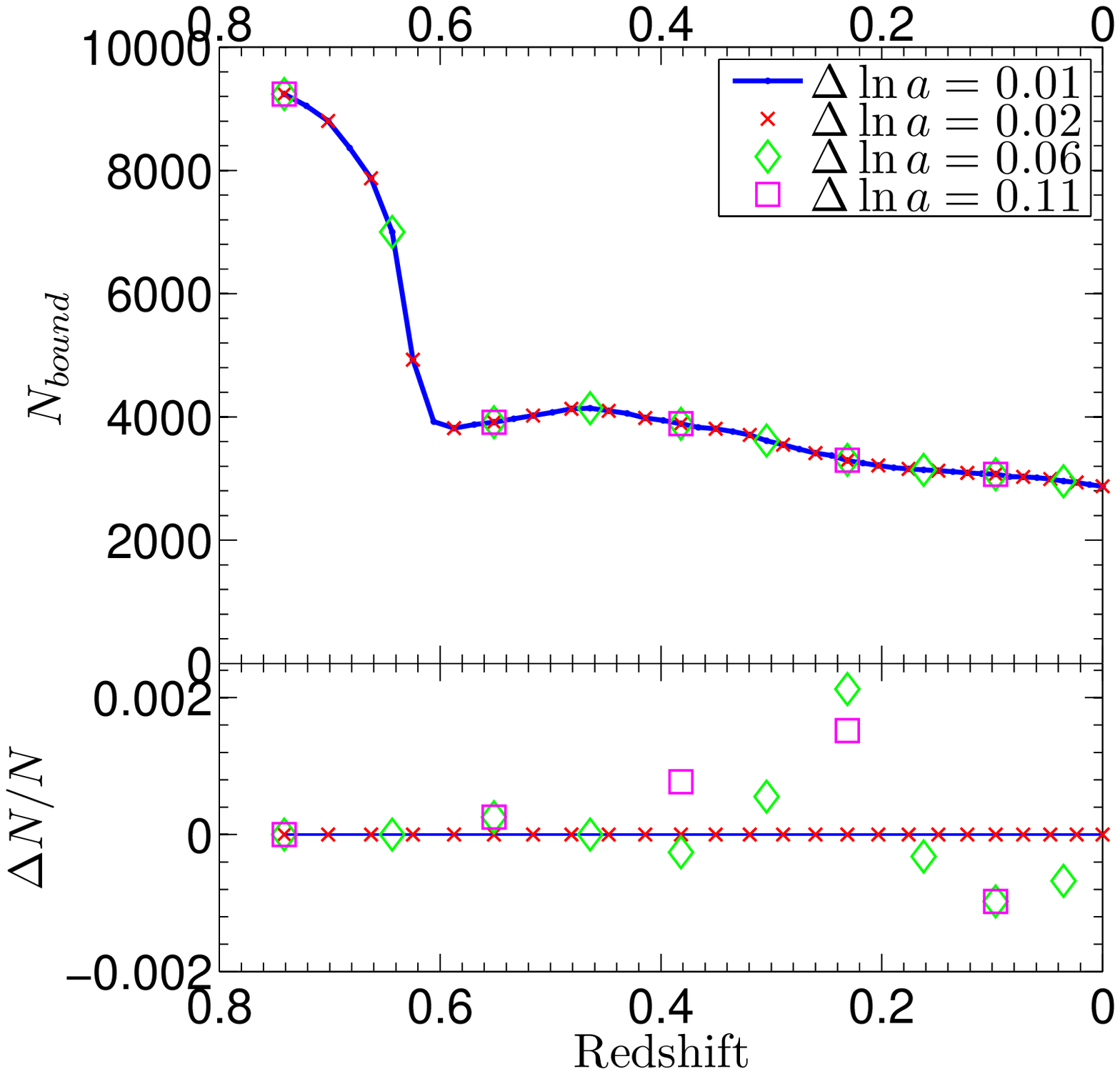}
\caption{Effect of time resolution on the self-bound mass. We apply
  the code skipping different number of snapshots at every tracing
  step, to test the stability of our unbinding algorithm with respect
  to time resolution. In the top panel, the solid line denotes for the
  bound mass when using all the snapshots available, while different
  symbols give the results of skipping different numbers of snapshots,
  labelled as different steps in the scale factor $a$. In the lower
  panel, we show the relative difference with respect to the highest
  resolution case.}\label{f:jump}
\end{figure}

The other aspect of the time resolution problem is whether the
unbinding routine would be dependent on the time resolution
adopted. We still use the halo S51G86 studied before to test this. As
shown in Figure~\ref{f:jump}, applying the adaptive unbinding
algorithm with different time resolution has no noticeable effect on
the resulting self-bound mass. This again demonstrates the robustness of 
our core-averaged unbinding algorithm in conjugation with our adaptive source
management.

\section{Unilaterally-matched subhaloes}\label{sec_uni}
Unilateral subhaloes in Figure~\ref{f:6702match} come in several situations. First there are
subhaloes found by one finder but missed by the other, and thus they
appear as part of a larger subhalo, either central or satellite. These
are reflected as vertical lines connecting up or horizontal lines
connecting right to bilateral subhaloes. Most unilateral subhaloes
belong to this case, with the majority having their correspondences
being the central. For those unilateral HBT subhaloes linked to a
satellite, most of them are in fact overlapping with their
correspondence, having a separation between their centres being the
order of the smoothing length of dark matter particles. In fact if
they can keep this overlapping state, these two subhaloes should be
treated as one, as merger between them has completed. However, because
overlapping subhaloes are still rare, and because it is not clear
whether these pairs have mixed completely, we do not integrate them
together in the current version of HBT.  We also notice that one
SUBFIND subhalo with $2\times10^3$ particles is missed by HBT,
indicating the tracing algorithm is still not perfect although it is
already massively optimized. However, as statistically found in our comparison using the cosmological simulation,
this kind of missed subhaloes can be as large as $10^4$ particles for SUBFIND, amounting to as high as 10 percent in number near that mass, while HBT only misses less than 1 percent for the highest missed mass of $10^3$ particles.

A second situation is when one subhalo finder only identifies the most
inner part of a subhalo and assigns a majority of surrounding bound
particles to the central subhalo, while its correspondence identifies
both parts. When doing the match, the surrounding particles from the
correspondence outweighs the inner part, linking the correspondence to
the central subhalo. This situation can be found as two adjacent
perpendicular lines connecting two unilateral subhaloes and the
central. It happens for both finders. In the SUBFIND case, it is due to
the truncation around the radius of equality as illustrated in
section~\ref{ss:profile} while in the HBT case we attribute this to the
effect of local accretion discussed in Appendix~\ref{s:acc}.

 The third situation is similar to the second one but happens for
 binary subhaloes, which has been discussed in the main text as ``sub-in-sub" switch.
 
\section{Accuracy of two estimators for the subhalo mass
function}\label{s:estimator}
\begin{figure}
\includegraphics[width=0.5\textwidth]{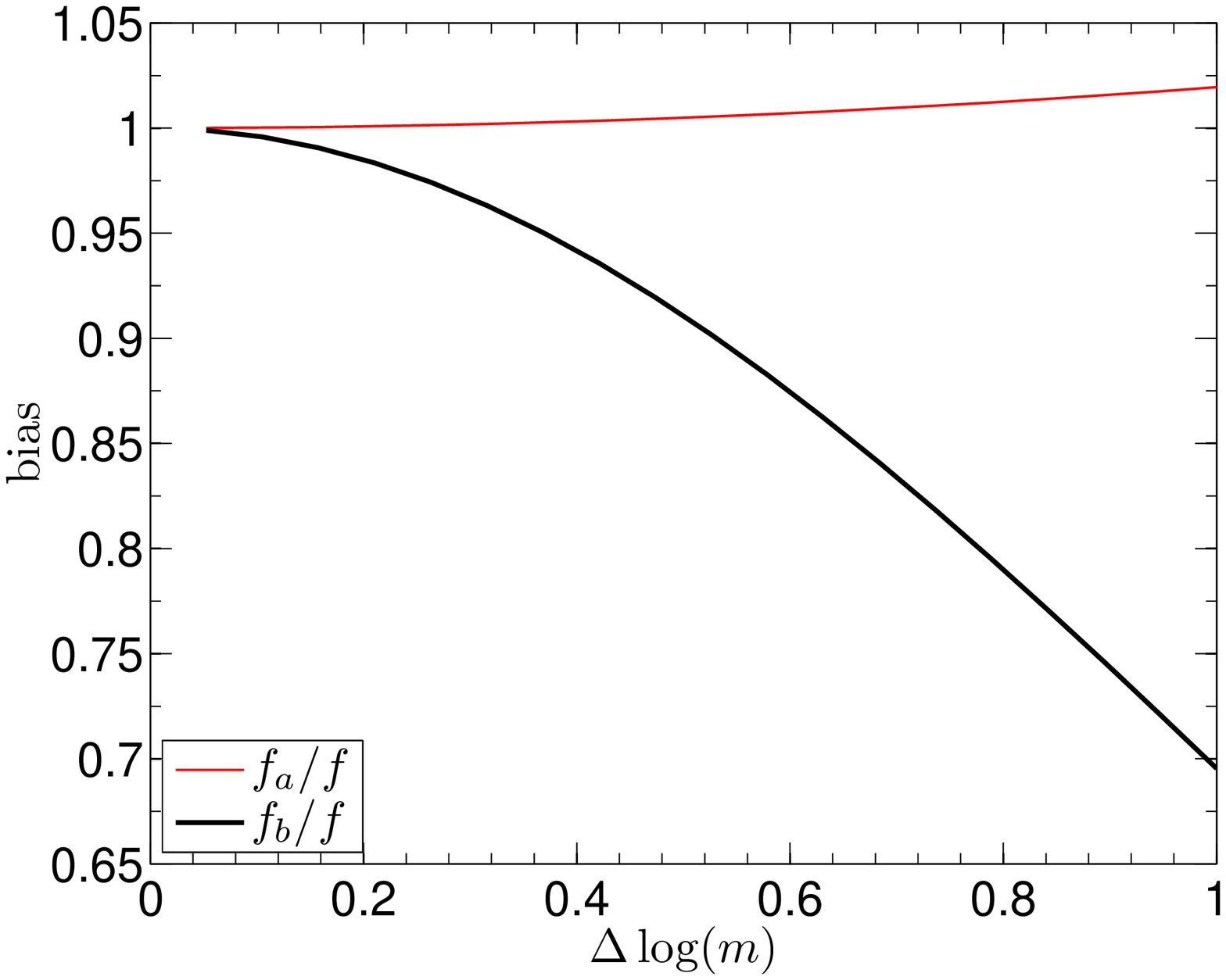}
\caption{Bias of mass function estimators as a function of mass bin size
$\Delta\log(m)=\log(k).$}\label{f:estimator}
\end{figure}

The subhalo mass function $f(m)\equiv dN/d\ln m$ where $m$ is subhalo
mass (either $M_{sub}$ or $M_{sub}/M_{host}$) can be estimated
directly from two discrete estimators as
\begin{equation}
f_a(\bar{m})=\frac{\Delta N}{\Delta\ln m}
\end{equation}
or
\begin{equation}
f_b(\bar{m})=\frac{1}{\Delta m}\sum_{m_1<m_i<m_2}{m_i}
\end{equation}
where $m_1$ and $m_2$ are the lower and upper limits of subhalo mass
bin, $\Delta N$ is the number of subhaloes in the bin and $\bar{m}$ is the
average subhalo mass in the bin.  In the limit $\Delta m\rightarrow
0$, we have $\bar{m}=m$ and $f_a(\bar{m})=f_b(\bar{m})=f(m)$. For a
finite mass bin, suppose $f(m)=m^{-1}$ ignoring the normalization,
then
\[
f_a(\bar{m})=-\frac{\Delta\frac{1}{m}}{\Delta\ln{m}}
\]\[
f_b(\bar{m})=\frac{\Delta\ln{m}}{\Delta{m}}
\]\[
f(\bar{m})=-\frac{\Delta\frac{1}{m}}{\Delta\ln{m}}
\]
It is ready to see that the estimator $f_a$ recovers exactly $f$.  For
a more general form $f(m)=m^{-\gamma}$, let $k=m_2/m_1$, then the
biases of the two estimators are given as
\begin{equation}
f_a(\bar{m})/f(\bar{m})=\frac{1}{-\gamma}\frac{k^{-\gamma}-1}{\ln{k}}(\frac{\gamma}{\gamma-1}\frac{k^{1-\gamma}-1}{k^{-\gamma}-1})^{\gamma}
\end{equation}
\begin{equation}
f_b(\bar{m})/f(\bar{m})=\frac{1}{1-\gamma}\frac{k^{1-\gamma}-1}{k-1}(\frac{\gamma}{\gamma-1}\frac{k^{1-\gamma}-1}{k^{-\gamma}-1})^{\gamma}
\end{equation}
Since the biases do not depend on $m$, the slope of the mass function
is recovered for both estimators while the normalizations differ. In
Figure~\ref{f:estimator} we show the bias of these two estimators for
$\gamma=0.9$. With bin size $\Delta\log(m)=0.5$, $f_b$ will
underestimate the mass function by 10 percent. Throughout this work we
use $f_a$ to calculate subhalo mass function except in
Figure~\ref{f:sat_acc} and Figure~\ref{f:mfun_resim} where $f_b$ is
used to give the total subhalo mass contained in each mass bin.

\end{document}